\documentclass[aps,prb,twocolumn,superscriptaddress,showkeys]{revtex4-1}
\pdfoutput=1

\usepackage{graphicx}
\usepackage[colorlinks=true,linkcolor=blue,citecolor=blue]{hyperref}

\begin{document}


\title{Fork stamping of pristine carbon nanotubes onto ferromagnetic contacts for spin-valve devices}

\author{J. Gramich}
\email[]{joerg.gramich@unibas.ch}
\author{A. Baumgartner}
\affiliation{Department of Physics, University of Basel, Klingelbergstrasse 82, 4056 Basel, Switzerland}
\author{M. Muoth}
\author{C. Hierold}
\affiliation{Department of Mechanical and Process Engineering, ETH Zurich, Tannenstrasse 3, 8092 Zurich, Switzerland}
\author{C. Sch{\"o}nenberger}
\affiliation{Department of Physics, University of Basel, Klingelbergstrasse 82, 4056 Basel, Switzerland}

\date{\today}

\begin{abstract}
We present a fabrication scheme called `fork stamping' optimized for the dry transfer of individual pristine carbon nanotubes (CNTs) onto ferromagnetic contact electrodes fabricated by standard lithography. We demonstrate the detailed recipes for a residue-free device fabrication and in-situ current annealing on suspended CNT spin-valve devices with ferromagnetic Permalloy (Py) contacts and report preliminary transport characterization and magnetoresistance experiments at cryogenic temperatures. This scheme can directly be used to implement more complex device structures, including multiple gates or superconducting contacts.
\end{abstract}


\maketitle


\section{Introduction}
Recent years have seen a tremendous increase of research activities based on clean fabrication schemes for carbon nanotube (CNT) quantum dot (QD) devices \cite{Cao:2005,Kuemmeth:2008,Steele:2009,Wu:2010,Muoth:2012,Pei:2012,Waissman:2013,Jung:2013,Viennot:2014,Ranjan:2015}, enabling the investigation of a variety of fundamental physical phenomena, including tunable QDs in the few electron regime \cite{Steele:2009,Pei:2012,Waissman:2013,Jung:2013,Ranjan:2015}, Fabry-Perot interference \cite{Jung:2013,Cao:2004}, spin-orbit interaction \cite{Kuemmeth:2008,Steele:2013}, valley spin-qubits \cite{Laird:2013} or the interaction between electron tunneling and the mechanical motion of the CNT \cite{Steele2:2009,Benyamini:2014}. All these discoveries, conducted by transport experiments in ultra-clean, suspended CNT QD systems, were made possible by novel fabrication schemes with pristine, as-grown CNTs that are never exposed to an electron beam (deposition of amorphous carbon), resists or solvents, which are believed to contaminate interfaces and the active structure \cite{Cao:2005,Wu:2010}. However, the combination of ultra-clean suspended CNTs with superconductors or ferromagnets has not been achieved, yet. On substrates and with standard processing, many new effects were reported recently for hybrid-CNT devices with superconducting or ferromagnetic contacts, including Cooper pair splitting (CPS) \cite{Herrmann:2010,Schindele:2012}, a possible source of spin-entangled electrons, the observation of Andreev bound states \cite{Pillet:2010} or the realization of electrically tunable spin-valve signals \cite{Sahoo:2005}. Such devices combined with clean, suspended CNT QDs could be used to realize several recent theoretical proposals, including Hanle-type experiments on QDs with ferromagnetic contacts \cite{Braun:2005}, coupling phonons in suspended CNTs either to resonant Andreev tunneling \cite{Zhang:2012}, or to spin-polarized currents \cite{Stadler:2014}, or in an electronic Bell test \cite{Braunecker:2013}.

In this study, we merge an ultra-clean fabrication approach with a ferromagnetic spin-valve structure, suspending a pristine, as-grown CNT over the ferromagnetic contacts in the last fabrication step. We choose an approach based on the mechanical transfer of the CNT \cite{Wu:2010,Muoth:2012,Pei:2012,Waissman:2013,Ranjan:2015} termed fork stamping \cite{Muoth:2012,Ranjan:2015}. Compared to a final CNT growth process directly on predefined electrode structures \cite{Cao:2005,Kuemmeth:2008,Steele:2009}, we are not limited to temperature-resistant materials because the CNT growth is performed independently from the actual device structure on a separate `transfer chip' \cite{Waissman:2013}. This key advantage of fork stamping allows us to process the actual device structure on the `electrical circuit chip' \cite{Waissman:2013} with arbitrary electrode materials - including temperature sensitive superconductors and ferromagnets - similar to standard devices on substrate. Because the transfer can be optically monitored and controlled, it allows a precise alignment and deterministic transfer of individual CNTs \cite{Muoth:2012}, in contrast to the more direct approach of Ref. \citenum{Viennot:2014}. To demonstrate our technique, we describe the fabrication of a spin-valve structure with ferromagnetic electrodes. We choose Permalloy (Ni$_{80}$/Fe$_{20}$, Py) as contact material which allows to accurately control the easy axis of the magnetization and the respective coercive field by shape anisotropy \cite{Aurich:2010}. We show that single CNTs can be mechanically transferred on top of ferromagnetic contacts, obtaining contact resistances comparable to values published in the literature and allowing magnetoresistance measurements in the QD regime at low temperatures. 

\section{Fork stamping of pristine CNTs onto arbitrary materials}
\begin{figure}[htb]
\includegraphics*[width=\linewidth]{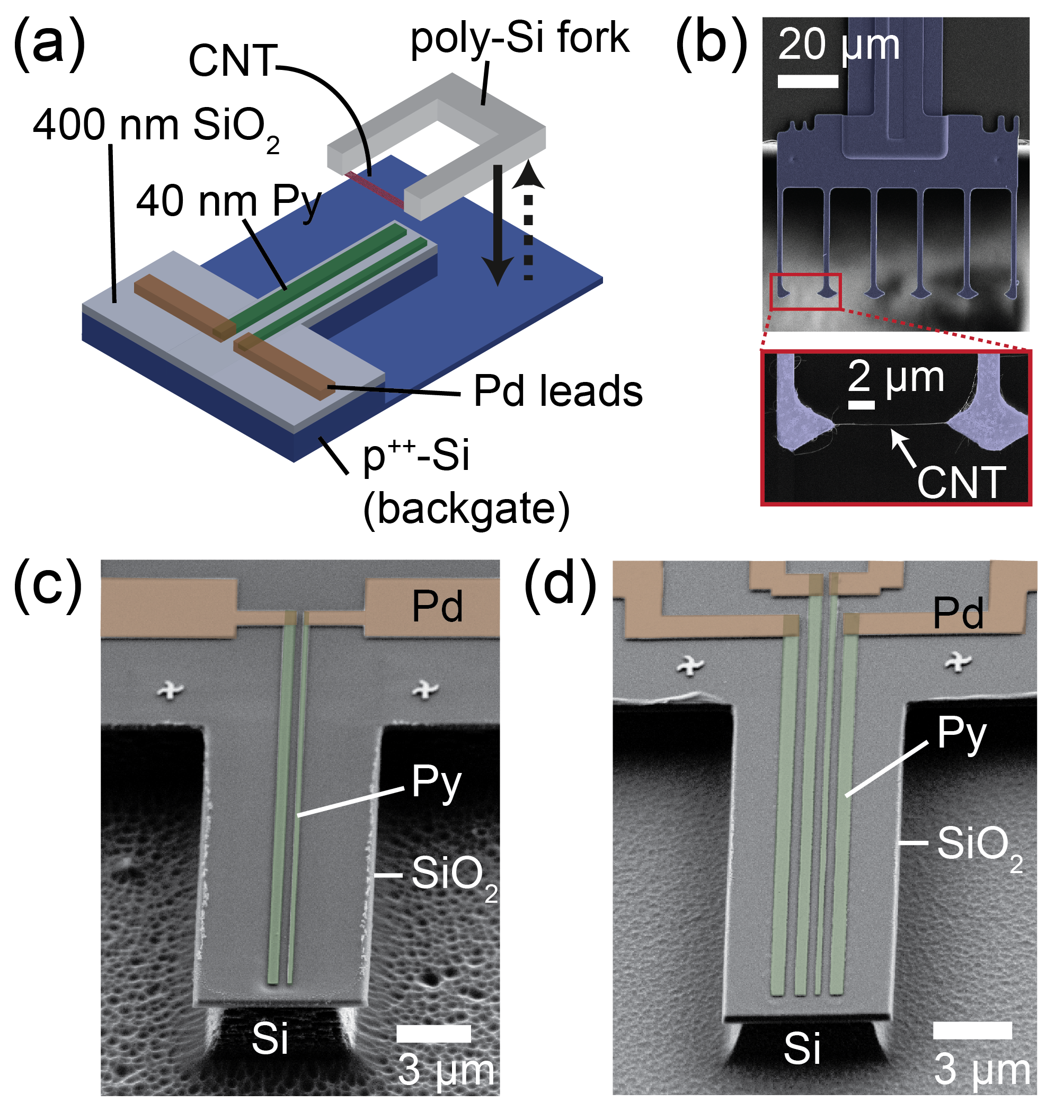}
\caption{\label{Fig:StampingPrinciple}Principle of fork stamping and individual components for the dry transfer. (a) Schematic of the sample layout and CNT transfer onto the electrical circuit chip with mesa structure. (b) False colored scanning electron microscopy (SEM) image of a moveable poly-Si fork (blue) at the edge of the transfer chip. Individual CNTs are grown across the arms of the fork (zoom-in). (c, d)  Tilted SEM images of electrical circuit chips with $6.8\,\mathrm{\mu m}$ wide, $3\,\mathrm{\mu m}$ high SiO$_2$/Si mesa structure, $40\,$nm thick ferromagnetic Py electrodes (light green) and connecting Pd leads (brown). The 4-terminal structure allows to electrically cut CNTs between the outer electrode pairs in the pushed-down state of the poly-Si fork.}
\end{figure}

Figure \ref{Fig:StampingPrinciple} (a) summarizes the dry transfer of pristine CNTs onto prefabricated `electrical circuit chips'. Following Refs. \citenum{Wu:2010,Muoth:2012,Pei:2012,Waissman:2013}, the electrical circuit preparation is detached from the growth of the CNTs, done separately on a `transfer chip'. We transfer the CNTs only in the last fabrication step under a light microscope and ambient conditions in a process called fork stamping, which allows the precise placement of a single CNT onto a suited, predefined electrical circuit \cite{Muoth:2012,Ranjan:2015}. Figure \ref{Fig:StampingPrinciple} (c) shows our electrical circuit chip consisting of a mesa structure with two ferromagnetic Py source-drain (SD) electrodes, a heavily p-doped Silicon (Si) wafer acting as backgate and Palladium (Pd) leads. As schematically depicted in Fig. \ref{Fig:StampingPrinciple} (a), a moveable Si fork structure, with ideally a single CNT grown across it, is pushed down on the predefined mesa structure (black arrow) while at the same time monitoring the current through the ferromagnetic SD contacts. Once contact has been detected and established, the fork is retracted (dashed arrow), ideally leaving an individual pristine CNT suspended over the ferromagnetic contacts. To immediately characterize the CNT device, we measure the differential conductance $G$ as a function of the backgate voltage. Undesired CNTs can then be removed by applying large SD bias voltages and the stamping procedure is repeated until an optimal CNT is found.

\paragraph{CNT growth on optimized transfer chip with retractable poly-Si forks}
CNTs are grown separately from the electrical circuit on a transfer chip with polycrystalline silicon fork structures, following previously reported protocols \cite{Muoth:2012,Muoth:2010}. Figure \ref{Fig:StampingPrinciple} (b) shows a scanning electron microscopy (SEM) image of such a fork structure protruding beyond the edge of the transfer chip. Each fork consists of multiple, $2\,\mathrm{\mu m}$ wide, $1.5\,\mathrm{\mu m}$ thick  poly-Si arms and can be retracted beyond the wafer edge, allowing to use different forks on the same transfer chip in one transfer session. CNTs are grown via chemical vapor deposition across the $8\,\mathrm{\mu m}$ wide gaps between the fork arms by using iron-loaded ferritin proteins as catalyst precursors \cite{Muoth:2010,Durrer:2009}. At best, one obtains maximally one individual CNT spanning each gap between the fork arms as visible in the zoom-in SEM image of Fig. \ref{Fig:StampingPrinciple} (b), for which the catalyst concentration has to be optimized on reference forks \cite{Muoth:2012}. Electron microscopy is omitted on the actual transfer chips to avoid carbon deposition, and to maintain the pristine, as-grown character of the transferred CNTs.

\paragraph{Fabrication of mesa structure and electrodes on electrical circuit chip}
\begin{figure*}[htb]%
\includegraphics*[width=\textwidth]{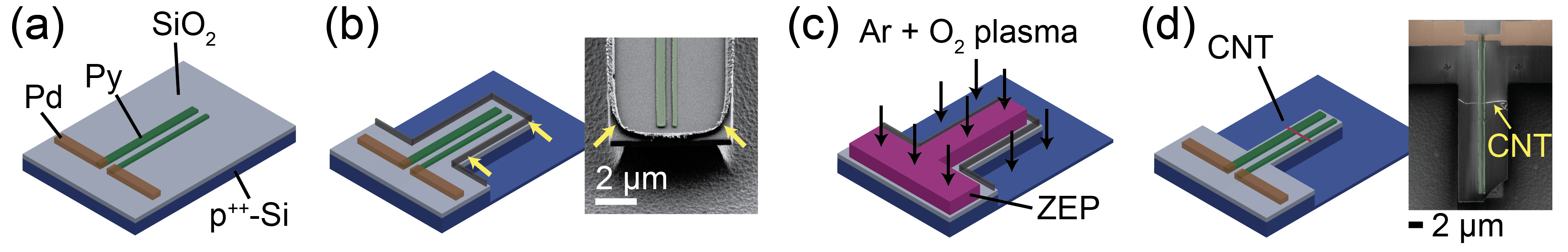}
\caption{\label{Fig:FabricationWorkflow}Schematic of the mesa fabrication workflow. (a) $40\,$nm thick, $160$/$380\,$nm wide ferromagnetic Py electrodes with connecting Pd leads are prepatterned in two consecutive e-beam lithography steps on top of a Si/SiO$_2$ wafer. (b) After the CHF$_3$/SF$_6$ mesa etching process and the removal of the PMMA resist mask, hard carbon-fluor polymer residues remain near the resulting mesa edge (arrows in the sketch and SEM image). (c) These undesired etching residues can be completely removed by Ar/O$_2$ plasma etching, where a patterned ZEP resist mask on the mesa structure exposes only areas with residues to the plasma. (d) Finished device (schematic and SEM image) after CNT transfer and electrical measurements.}
\end{figure*}
The electrical circuit chip with the mesa structure and the ferromagnetic contacts is fabricated independently of the transfer chip in a five-step electron-beam lithography (EBL) process (Fig. \ref{Fig:FabricationWorkflow}). We use a heavily p-doped Si wafer with a $400\,$nm thick thermal oxide top layer as substrate. After the deposition of Au markers, two $20\,\mathrm{\mu m}$ long, $40\,$nm thin ferromagnetic Py electrodes of variable width are patterned with a pitch of $400\,$nm by a previously optimized EBL step with ZEP520A resist, thermal electron gun evaporation and lift-off process \cite{Samm:2014}. The use of thin, high aspect ratio Py contacts allows us to obtain single-domain contacts with the magnetic easy axis along the strip direction and accurately control the coercive or switching fields by the width of the contacts \cite{Aurich:2010}. Here, we choose $160\,$nm and $380\,$nm wide Py strips to obtain well separated switching fields of the electrodes \cite{Samm:2014}. The Py strips are contacted by $50\,$nm thick Pd leads in another EBL step (see Fig. \ref{Fig:FabricationWorkflow} (a)). This results in a geometry suited for spin-valve experiments in the quantum dot (QD) regime at low temperatures \cite{Sahoo:2005,Aurich:2010,Samm:2014}, where the heavily p-doped Si wafer can be used as a backgate to tune the chemical potential of the QD. For creating the mesa structure in the SiO$_2$/Si, we first spin-coat a $1.2\,\mathrm{\mu}$m thick PMMA 950k resist layer as etching mask. This resist is exposed and developed by EBL, leaving the mesa structure and Py electrodes protected. For etching into the SiO$_2$/Si around the mesa structure, we use an optimized, anisotropic and selective reactive ion etching (RIE) process \cite{Legtenberg:1995} in an Oxford Plasmalab 80 Plus system, resulting in a T-shaped etch profile (compare Fig. \ref{Fig:StampingPrinciple} (c)). First, a $10\,$min. long CHF$_3$ RIE etching is used for the removal of the $400\,$nm thick SiO$_2$ layer with a flow of $8\,$sccm CHF$_3$, pressure $p=50\,$mTorr and power $200\,$W, leading to an etching selectivity of $\sim 2.5$ compared to the resist mask. A subsequent $2.5\,$min. anisotropic and selective SF$_6$/O$_2$ RIE with a flow of $13\,$sccm SF$_6$ and $5\,$sccm O$_2$, $p=75\,$mTorr, power $100\,$W and an etching selectivity of $\sim 7$ compared to the resist creates a $3\,\mathrm{\mu}$m deep trench in the p-doped Si. The anisotropic profile stems from a Si sidewall passivation by oxygen species during the etching process \cite{Legtenberg:1995}, which also leads to the observed porous structure in the Si. Before lift-off, the resist mask is typically etched down to a thickness of $300\,$nm in an Ar/O$_2$ RIE step to minimize residues from the etch process. After the lift-off procedure in warm N-Methyl-2-pyrrolidone (NMP) or acetone, we typically encounter large carbon-fluor polymer residues close to the edge of the mesa structure (arrows in Fig. \ref{Fig:FabricationWorkflow} (b)) which severely hinder a successful CNT transfer. These residues originate from polymerization and passivation layers formed in the CHF$_3$/SF$_6$ RIE etching on the side of the resist mask \cite{Kim:2006} and are not removable by the solvents in the lift-off process. We found that the selectivity of the etch process is optimal only in a narrow parameter range and cannot be optimized simultaneously to protect the mesa without residues. Sonication to remove the residues mechanically, as used in Ref. \citenum{Ranjan:2015}, typically leads to a partial collapse of the mesa structure and affects the sensitive Py strips. We remove these polymer residues using a 5th EBL process: Spincoating another $450\,$nm thick ZEP520A resist layer on the wafer covers the entire mesa structure with the Py strips and the residues. We only expose and develop areas close to the edge of the mesa structure in another EBL step, resulting in a mask open only in the area of the residues, as sketched in Fig. \ref{Fig:FabricationWorkflow} (c). The residues can thus be reproducibly removed in a standard Ar/O$_2$ plasma without exposing the Py strips to O$_2$ plasma. After another lift-off procedure this results in clean, $6.8\,\mu$m wide and $3\,\mathrm{\mu}$m high mesa structures etched into the SiO$_2$/Si substrate, which allow access for the CNT fork stamping. Figure \ref{Fig:StampingPrinciple} (c) and (d) show two different realized geometries, one with two ferromagnetic contact electrodes, the other with four. While in the first geometry only a mechanical transfer of the CNTs can be implemented, the 4-terminal geometry can also be used for an electrical cutting of the CNT described in detail later, or, alternatively, for non-local spin experiments similar to graphene \cite{Tombros:2007}. The comparatively wide mesa relative to the $8\,\mathrm{\mu m}$ fork gap improves the CNT transfer, while the T-shaped underetched mesa profile is beneficial for avoiding CNT-induced electrical shorts from the contacts to the p-doped backgate. To monitor the current through the SD contacts during CNT transfer, all electrical circuit chips are glued into chip carriers and wire-bonded prior to the transfer of CNTs. Figure \ref{Fig:FabricationWorkflow} (d) shows a schematic and SEM image of a device after successful CNT transfer, demonstrating that single CNTs (yellow arrow) can be transferred onto ferromagnetic contacts.

\paragraph{CNT transfer}
In the final fabrication step, the CNTs are transferred from the fork structures onto the electrical circuit chip \cite{Muoth:2012}. Immediately prior to the CNT transfer, the electrical circuit chips with the ferromagnetic Py electrodes have to be cleaned from surface oxides in a $25\,$s long Argon plasma etch. This step is crucial, since without it no electrical contact is formed due to contamination and oxidation of the electrodes. The transfer is done using a micro-manipulator setup under ambient conditions. To reduce further oxidation, a nitrogen flow is applied around the electrical circuit chip. The transfer chip - with the CNT forks protruding beyond the edge of the wafer - is carefully mounted on a three axis piezo controlled transfer arm. Next, the forks with CNTs are carefully aligned with the mesa structure using an optical microscope. Finally, the fork is pressed down over the mesa structure and the contacts (black arrow in Fig. \ref{Fig:StampingPrinciple} (a)). Simultaneously, we monitor the current through two SD terminals in a voltage-biased setup. In contrast to previous experiments with normal Pd or Au electrodes \cite{Muoth:2012,Waissman:2013,Ranjan:2015}, we have to apply relatively large SD voltages of $1-3\,$V to electrically register every CNT `touch-down' event when a CNT bridges the SD contacts, carefully studied by SEM on reference samples following the transfer. When a successful contact is detected, there are two ways to deposit the CNT. First, it can be mechanically torn off by keeping the fork pushed down and retracting it parallel to the chip surface using the built-in mechanism, leaving the CNT in place and suspended over the SD contacts due to van-der-Waals forces. Second, in a 4-terminal device (Fig. \ref{Fig:StampingPrinciple} (d)), the CNT can be selectively cut between the outer pair of contacts using large electrical currents, following the approach of Ref. \citenum{Waissman:2013}. In the pushed-down state of the transfer fork, the application of a large voltage between the two outer pairs of contacts (while maintaining the inner ones on the same potential) breaks the CNT at a single point between each of the two outer contact pairs due to Joule heating \cite{Waissman:2013}. The transfer fork can then be lifted, leaving only a part of the CNT suspended over the inner contacts. Typical cutting currents for single wall CNTs are on the order of $15-25\,\mathrm{\mu A}$, consistent with Ref. \citenum{Waissman:2013}. Usually, we obtain room temperature device resistances of $R_\mathrm{SD} \sim 1-10\,\mathrm{M\Omega}$ for a single CNT. The contacts can already be annealed during stamping by driving the SD voltage up and down, a process which will be described in detail in the next section. In case that the resistance is too low ($R_\mathrm{SD} \ll 500\,\mathrm{k\Omega}$) - meaning that bundles or several CNTs were transferred, as inferred from SEM images of reference samples - the CNTs can be removed by applying a large SD bias voltage. In the cutting procedure for the 4-terminal geometry discussed above, this is immediately evident when one observes several steps in the monitored IV-curves or much higher currents than $25\,\mathrm{\mu A}$ are required to cut the CNT. After the removal of undesired CNTs, the electrical circuit chip can be used again for further transfers until an optimal CNT is found. Already during transfer, we observe that stamped CNT devices on ferromagnetic Py contacts degrade to a much higher device resistance on the timescale of a few minutes, often reaching values as high as $100\,\mathrm{M\Omega}$ when the device had an original resistance of $1\,\mathrm{M\Omega}$. This suggests that the surfaces of the Py contacts oxidize fast. The exposure of samples to air between the CNT transfer under nitrogen flow and the mounting of samples in our cryogenic measurement setup is thus reduced to a minimum.

\section{Electrical characterization}
\begin{figure}[htb]
\includegraphics*[width=\linewidth]{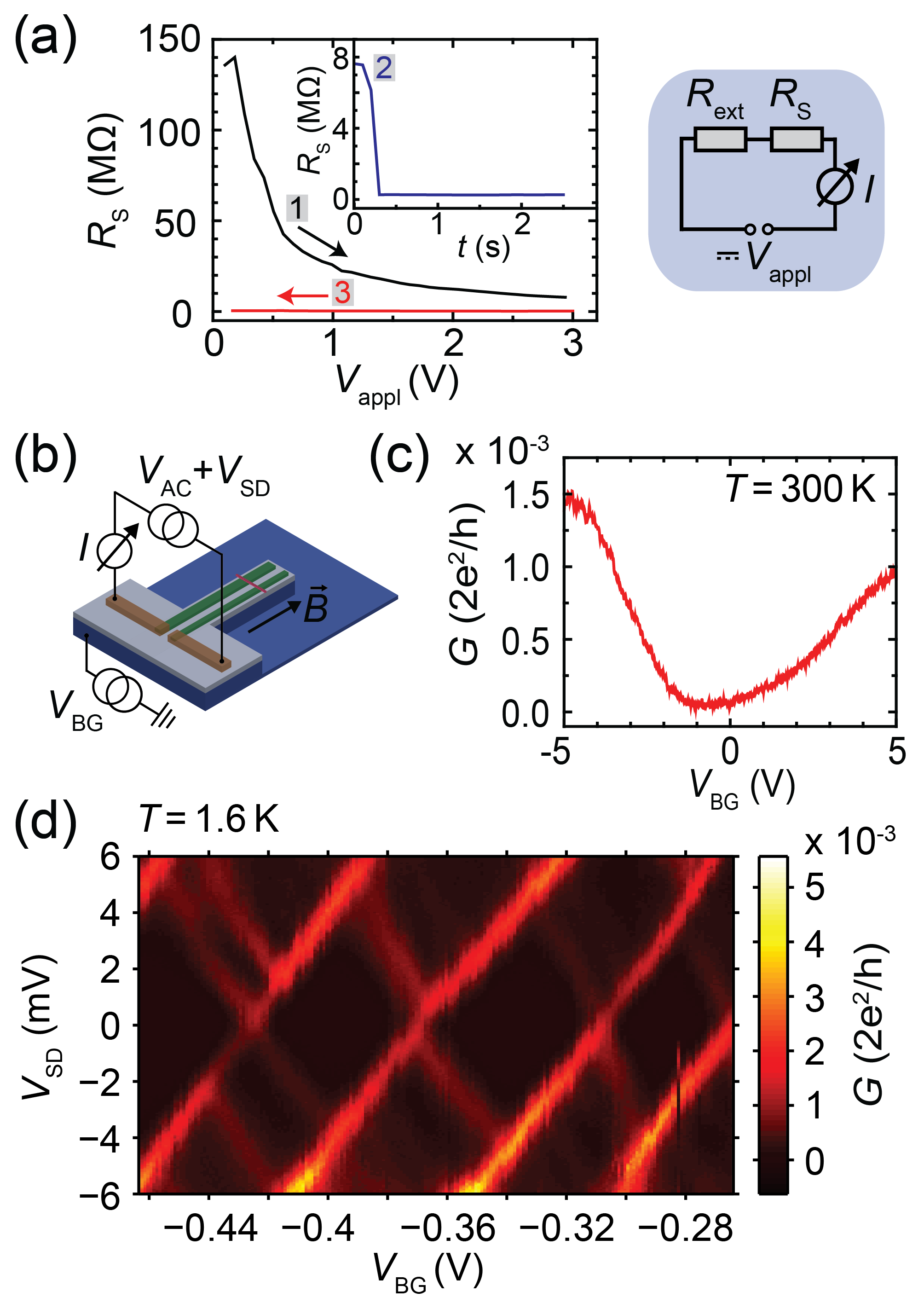}
\caption{\label{Fig:ElectricalCharacterization}Electrical characterization of stamped CNT devices with Py contacts. (a) Typical contact annealing trace, recorded at room temperature in low-pressure He-atmosphere with the annealing circuit depicted on the right. The sample resistance $R_\mathrm{S}$ is plotted as a function of the voltage $V_\mathrm{appl}$ applied over the sample and a pre-resistor $R_\mathrm{ext}=100\,\mathrm{k\Omega}$ for the up- (1) and downsweep (3). The inset shows the waiting trace (2) at $V_\mathrm{appl}=3\,$V. (b) Measurement scheme for electrical measurements. (c) Differential conductance $G$ as a function of backgate voltage $V_\mathrm{BG}$ for an annealed device at room temperature. (d) $G$ as a function of $V_\mathrm{BG}$ and source drain bias $V_\mathrm{SD}$ for an annealed device at $T=1.6\,$K.}
\end{figure}
The already wire-bonded CNT devices are built into a cryogenic variable temperature insert allowing measurements in the range of $T=1.5-300\,$K. Samples usually degrade and have a relatively high resistance compared to the values directly after CNT transfer. To measure a signal at low temperatures, the devices have first to be annealed in the cryostat at room temperature and in low-pressure He atmosphere. Fig. \ref{Fig:ElectricalCharacterization} (a) shows the electrical circuit for annealing and a typical `annealing trace'. A DC voltage $V_\mathrm{appl}$ is applied over an external resistor $R_\mathrm{ext}$ in series with the sample $R_\mathrm{S}$, measuring the current, from which $R_\mathrm{S}$ can be calculated. A typical annealing cycle consists of ramping the voltage up to some predefined value $V_\mathrm{max}$ (black trace 1), a waiting trace at $V_\mathrm{max}$ for a given time (blue trace 2) and the backtrace (red curve 3). This is repeated several times with increasing $V_\mathrm{max}$ until a device resistance change to reasonable values in the $<1\,\mathrm{M\Omega}$ range is observed. Figure \ref{Fig:ElectricalCharacterization} (a) shows the traces of a successful annealing, with a large device resistance at the start and a final device resistance of $R_\mathrm{S}=250\,\mathrm{k\Omega}$ originating from the abrupt change in the waiting trace (2) at $V_\mathrm{max}=3\,$V (inset). In contrast to the current annealing usually applied for suspended graphene devices where the cleaning and device changes are ascribed to Joule heating \cite{Moser:2007}, it is evident that such large resistance changes can only be caused by the contact resistance, showing that the annealing is actually not only cleaning the CNT, but mainly a contact resistance change. We note that the currents through the device prior to the resistance change are in the sub-$\mathrm{\mu}$A regime. These characteristics are similar for most samples with device resistance changes often occurring after a certain waiting time at $V_\mathrm{max}$ and agree very well with previous findings on surface-oxidized Pd/PdO contacts to CNTs \cite{Jones:2006}. We speculate that a large voltage portion drops across the contact interfaces, leading to an irreversible dielectric breakdown in the oxide barrier on the Py surface, possibly creating permanent percolation paths to the CNT \cite{Jones:2006,Lombardo:2005}. This results in a low impedance CNT device with relatively transparent contacts. For further electrical characterization, we use the measurement setup depicted in Fig. \ref{Fig:ElectricalCharacterization} (b). We measure the differential conductance $G=\mathrm{d}I/\mathrm{d}V$ using standard lock-in amplifiers as a function of the backgate voltage $V_\mathrm{BG}$ or source drain bias $V_\mathrm{SD}$ in a spin-valve geometry. To characterize the metallic or semiconducting nature of the stamped CNT, we measure $G$ at room temperature as function of $V_\mathrm{BG}$, seen in Fig. \ref{Fig:ElectricalCharacterization} (c) for a semiconducting CNT. Figure \ref{Fig:ElectricalCharacterization} (d) shows the charge stability diagram of such a device at $T=1.6\,$K, where we measure $G$ as a function of backgate voltage and the bias. Clear Coulomb Blockade (CB) diamonds are visible, indicating that a single QD forms in the CNT suspended over the two SD contacts. From the measurements, we can extract a  backgate leverarm of $\eta \sim 0.082\,$eV/V only slightly smaller than for CNTs on substrate, a charging energy $U_\mathrm{c}\sim 4\,$meV and from the excited states a level spacing of $\delta E=1.9\,$meV.  From the extracted level spacing, we can roughly estimate an effective QD size of $L = 1\,\mathrm{\mu m}/\delta E\,\mathrm{(meV)} = 0.52\,\mathrm{\mu m}$ \cite{Nygard:1999}, in reasonable agreement with the designed contact pitch and center-to-center separation of $400\,$nm, respectively $670\,$nm.

\section{Magnetoresistance}
\begin{figure}[htb]
\includegraphics*[width=\linewidth]{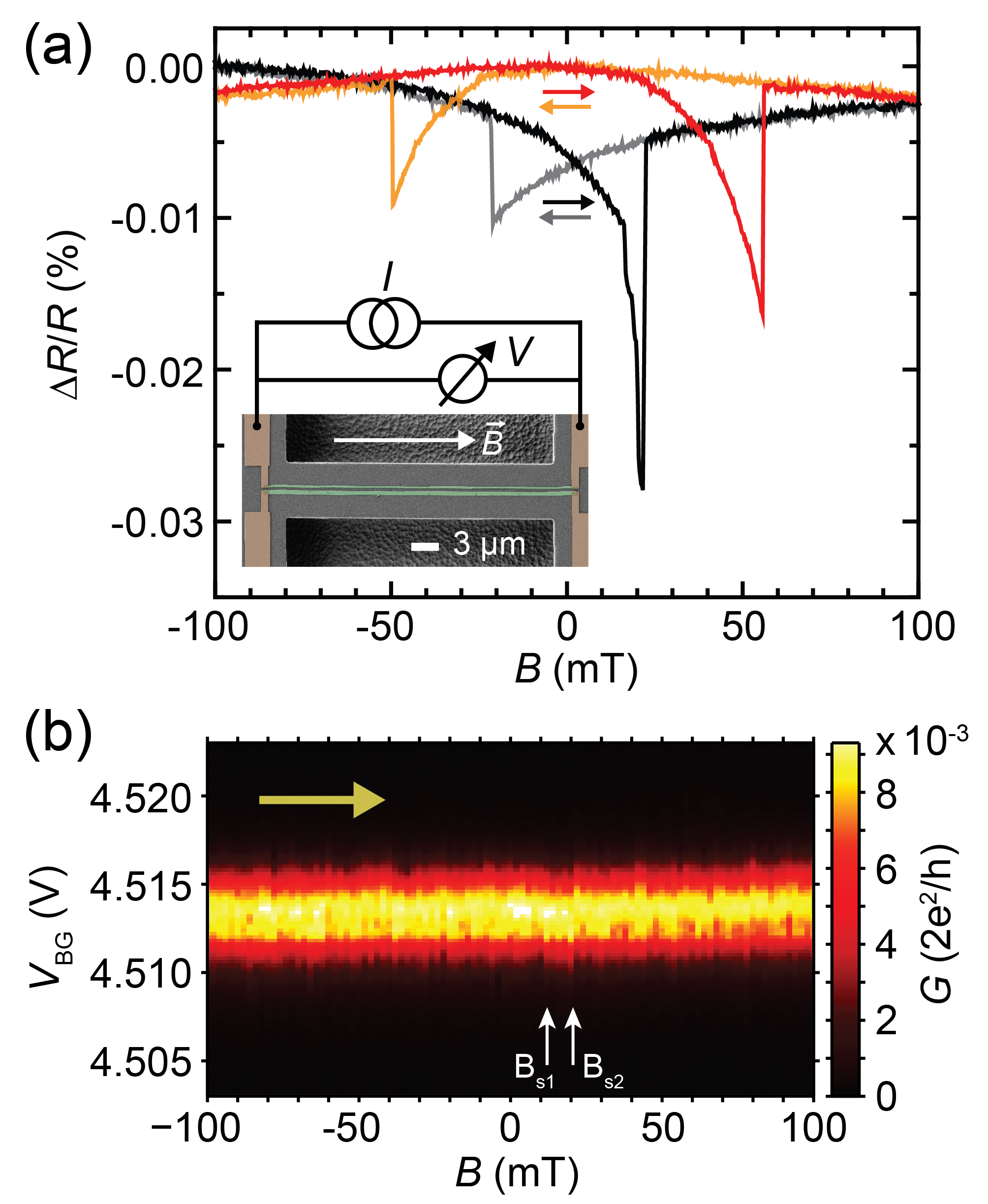}
\caption{\label{Fig:SpinSignals}Magnetoresistance measurements at $T=1.6\,$K. (a) Anisotropic magnetoresistance (AMR) of individual, $35\,\mathrm{\mu m}$ long, $160$/$380\,$nm wide (red and orange/black and grey curve) Py strips on a mesa structure. Small horizontal arrows denote the color-coded sweep direction of $B$. Inset: measurement setup and sample geometry with Py strips on a mesa bridge. (b) Differential conductance $G$ of a CNT device as function of the backgate voltage $V_\mathrm{BG}$ and an external magnetic field $B$ applied in parallel to the Py strips. The big horizontal arrow labels the magnetic field sweep direction, small vertical arrows denote the expected switching fields $B_\mathrm{s1}$ and $B_\mathrm{s2}$ for this device.}
\end{figure}
We perform local magnetoresistance measurements at $T=1.6\,$K in a standard QD spin-valve geometry. First, to characterize the quality of the ferromagnetic Py strips after processing, we perform anisotropic magnetoresistance (AMR) measurements following Refs. \citenum{Aurich:2010,Samm:2014}. For this, we use the sample geometry and measurement setup depicted in the inset of Fig. \ref{Fig:SpinSignals} (a). Here, the $35\,\mathrm{\mu}$m long Py strips reside on a mesa bridge structure surrounded by trenches in the  SiO$_2$/Si on two sides only, enabling transport measurements through each strip separately. It is possible to transfer CNTs also on this structure by tilting the transfer forks. We measure the resistance through individual Py strips while ramping the external magnetic field $B$ applied parallel to the strip axis. Figure \ref{Fig:SpinSignals} (a) shows the respective up- and downsweep (arrows) for a $160\,$ (red/orange) and $380\,$nm (black/grey) wide Py strip. Sharp resistance changes at $B_\mathrm{s1}=21\,$mT for the wide and $B_\mathrm{s2}=52\,$mT for the narrow Py strip indicate a sign reversal of the magnetization. These values agree very well with our previous strip characterization on flat substrates \cite{Aurich:2010,Samm:2014}. The sharp switching indicates that the bulk behavior of the magnetization remains intact after processing. To assess the magnetoresistance (MR) through the CNT, we measure MR maps over a single CB conductance maximum \cite{Samm:2014}, see Fig. \ref{Fig:SpinSignals} (b). Here, the conductance $G$ through the CNT is plotted as a function of the backgate voltage for the up-sweep of $B$ (large arrow). For this peculiar device with $380\,$ and $500\,$nm wide Py strips, we expect switching fields of $B_\mathrm{s1}=12-15\,$mT \cite{Samm:2014} and $B_\mathrm{s2}=21\,$mT, indicated by vertical arrows in Fig. \ref{Fig:SpinSignals} (b). A small change in amplitude and position of the CB conductance maximum might be visible at these positions, but the data remain inclusive for a reliable interpretation. The MR signal of a spin-valve device is proportional to the conductance change between the parallel and antiparallel magnetization configuration of the two contacts. We would expect a MR signal of $\sim 10\%$ for Py \cite{Samm:2014}, but the conductance noise of the device coming from both amplitude and position fluctuations of the CB resonances is on the same order of magnitude or larger. Further repetitions of up- and downsweeps on the same and other devices show similar features also with different switching fields. The conductance noise and instability of the devices, also apparent in the charge stability diagram of Fig. \ref{Fig:ElectricalCharacterization} (d), are still too large to detect reliably any spin signals in the CNT spin-valves. 
We speculate that these fluctuations of the CB resonances are caused by instabilities in the contact interfaces. The annealing curves clearly show that the ferromagnetic contact interfaces are not completely stable due to oxidation and might have inherent charge traps in the oxide, possibly close to the contact area. The current difficulties in magnetoresistance experiments can be overcome by using an in-situ setup similar to the one used in Ref. \citenum{Waissman:2013}, using less oxidizing ferromagnetic materials as e.g. PdNi alloys or implementing oxygen tight tunnel barriers on top of the ferromagnetic contacts prior to a CNT transfer. Hexagonal boron nitride (hBN), for example, holds great promise for longer spin-life times in bottom-up fabricated CNT spin-valve devices, an approach currently followed intensively for graphene spin-valves \cite{Fu:2014,Droegeler:2014}. 

\section{Conclusions}
We report a fabrication scheme suited for the mechanical transfer of individual CNTs onto ferromagnetic contacts, leading to pristine, as-grown CNTs suspended over ferromagnetic electrodes in a spin-valve geometry. This fabrication scheme can be readily extended to other contact materials, including complicated device layouts with several gates and superconducting as well as ferromagnetic electrodes, enabling the combination of ultra-clean QD systems with hybrid-CNT devices. Using these recipes, we demonstrate that single CNTs can be contacted on top of ferromagnetic contacts with reasonable device resistances achieved by an electrical contact annealing, and that we are able to form single QDs suited for transport studies at low temperatures. No clear spin-valve signal could be found yet in magnetoresistance experiments at low temperatures, which we tentatively ascribe to interface properties resulting in an increased conductance noise.

\begin{acknowledgments}
We thank S. Ilani, A. Bachtold, V. Ranjan, G. Puebla-Hellmann and J. Samm for fruitful discussions. The CNT stamping was done in the FIRST lab of ETHZ. We acknowledge generous access to that facility. This work was financially supported by the Swiss National Science Foundation (SNF), the Swiss Nanoscience Institute (SNI), the Swiss NCCR QSIT, the ERC project QUEST and the EU FP7 project SE2ND.
\end{acknowledgments}

\bibliography{LitFerroStamping}

\end{document}